\documentclass[10pt]{article}
\usepackage[OE]{express}
\usepackage{gensymb}

\begin{document}
\title{Optical bandgap engineering in nonlinear silicon nitride waveguides}
\author{Clemens~J.~Kr\"uckel\authormark,{*} Attila F\"ul\"op, Zhichao Ye, Peter A. Andrekson, and Victor Torres-Company}
\address{Photonics Laboratory, Department of Microtechnology and Nanoscience, Chalmers University of Technology, SE-412 96 Gothenburg, Sweden}
\email{\authormark{*}kruckel@chalmers.se} %% email address is required



\begin{abstract}
\noindent Silicon nitride is a well-established material for photonic devices and integrated circuits. It displays a broad transparency window spanning from the visible to the mid-IR and waveguides can be manufactured with low losses. An absence of nonlinear multi-photon absorption in the erbium lightwave communications band has enabled various nonlinear optic applications in the past decade. Silicon nitride is a dielectric material whose optical and mechanical properties strongly depend on the deposition conditions. In particular, the optical bandgap can be modified with the gas flow ratio during low-pressure chemical vapor deposition (LPCVD). Here we show that this parameter can be controlled in a highly reproducible manner, providing an approach to synthesize the nonlinear Kerr coefficient of the material. This holistic empirical study provides relevant guidelines to optimize the properties of LPCVD silicon nitride waveguides for nonlinear optics applications that rely on the Kerr effect.
\end{abstract}

%\ocis{(130.3120) Integrated optics devices; (160.4330) Nonlinear optics materials; (190.4390) Nonlinear optics, integrated optics; (220.4241) Nanostructure fabrication.} 

%%%%%%%%%%%%%%%%%%%%%%% References %%%%%%%%%%%%%%%%%%%%%%%%%

%%%%%%%%%%%%%%%%%%%%%%%%%%  body  %%%%%%%%%%%%%%%%%%%%%%%%%%
%%%%%%%%%%%%%%%%%%%%%%%%%%%%%%%%%%%%%%%%%%%%%%%%%%%%%
%%
\section{Introduction}
%%
%%%%%%%%%%%%%%%%%%%%%%%%%%%%%%%%%%%%%%%%%%%%%%%%%%%%%

Silicon nitride is a well-established material in the microelectronics industry, where it has been used as electrical and thermal insulator in electric circuits. This material can be processed using the mature fabrication lines for electronic integration. Silicon nitride also features a set of optical properties that makes it an ideal choice for many applications that require integration of photonic devices on chip. Examples include ultra-high-Q resonators \cite{ultra-high-q} and filters based on ultra-low-loss waveguides \cite{bauters}. Silicon nitride can also be co-integrated with active components such as amplifiers \cite{amplifiers}, modulators and detectors \cite{detectors}. The broad transparency window extends towards the blue wavelength range, which allows for processing visible light on chip \cite{visible-baets,visible_light} and biophotonic sensing  applications \cite{triplex-sensing}.\\
Another important feature of silicon nitride is that in its stoichiometric form (Si$_3$N$_4$), it has a nonlinear Kerr coefficient ten times higher than silica \cite{ikeda}. Its large optical bandgap and high thermal damage threshold permit high optical intensities to build up in the system without suffering from nonlinear losses in the near infrared. The potential of this material for nonlinear optics has been demonstrated with the generation of microsesonator based frequency combs \cite{levy-comb,comb-kippenberg,comb-perdue}, octave-spanning supercontinuum \cite{supercontinuum-cornell,lionix-sc-gen} and wavelength conversion \cite{wl_conv_kruckel,wl_conv_srinivasan}.\\
The optical properties of silicon nitride depend on the conditions under which the material is deposited. There is a wealth of empirical studies that demonstrate that varying the flow ratio of the precursor gases during chemical vapor deposition modifies the stoichiometry of the film \cite{refractive_index_vs_siN2,LPCVD_si-nitride}. This results in a change of refractive index, optical bandgap and film stress \cite{refractive_index_vs_siN2,stress_vs_siN,stress_vs_siN2}. Increasing the relative content of silicon vs nitride yields a silicon-rich nitride film \cite{LPCVD_si-nitride}. This is highly relevant for nonlinear optics applications because waveguides fabricated from silicon rich nitride tend to have a higher nonlinear Kerr coefficient \cite{kruckel:Opt-Exp,PECVD-SiN,PECVD-SiN-singapore}. Lacava et al. recently demonstrated that the Kerr coefficient can be enhanced by an order of magnitude in silicon nitride waveguides fabricated via plasma-enhanced chemical vapor deposition (PECVD) by varying the silicon content of the film \cite{PECVD-SiN}. However, most nonlinear optic applications are based on silicon nitride deposited via low-pressure chemical vapor deposition (LPCVD) \cite{lionix-sc-gen,levy-comb,si3n4_perdue,si3n4_epfl}. Despite the popularity of this platform, the impact of the film stoichiometry on the nonlinear properties of LPCVD silicon nitride is not well known.\\
In this paper, we focus on the full characterization of the nonlinear properties of LPCVD silicon nitride waveguides. We fabricated waveguides with five different compositions, from stoichiometric to the maximum silicon content that is allowed in our furnace. We show that the gas flow ratio modifies the optical properties of the material but in a highly reproducible fashion. We present a holistic empirical study, highlighting how the gas flow ratio can be used to engineer the optical properties of the waveguide for broadband nonlinear optic applications.\\
The remainder of this work is as follows. In section 2 we present the characterization of the optical properties of the film and the waveguide fabrication process. Section 3 covers the impact of the silicon nitride composition on waveguide propagation loss and group velocity dispersion engineering. In section 4 we focus on the evaluation of material dependent Kerr nonlinearities and assess the impact of waveguide confinement on the nonlinear Kerr parameter. In section 5 we compare the measured Kerr nonlinear coefficients with theoretical expectations and establish a comparison between LPCVD and PECVD silicon nitrides as well as other material platforms used for nonlinear optic applications.

%%%%%%%%%%%%%%%%%%%%%%%%%%%%%%%%%%%%%%%%%%%%%%%%%%%%%
%%
\section{Fabrication and material properties}
%%
%%%%%%%%%%%%%%%%%%%%%%%%%%%%%%%%%%%%%%%%%%%%%%%%%%%%%

\begin{figure}[!t]
  \begin{centering}
  \includegraphics[width=1.0\linewidth]{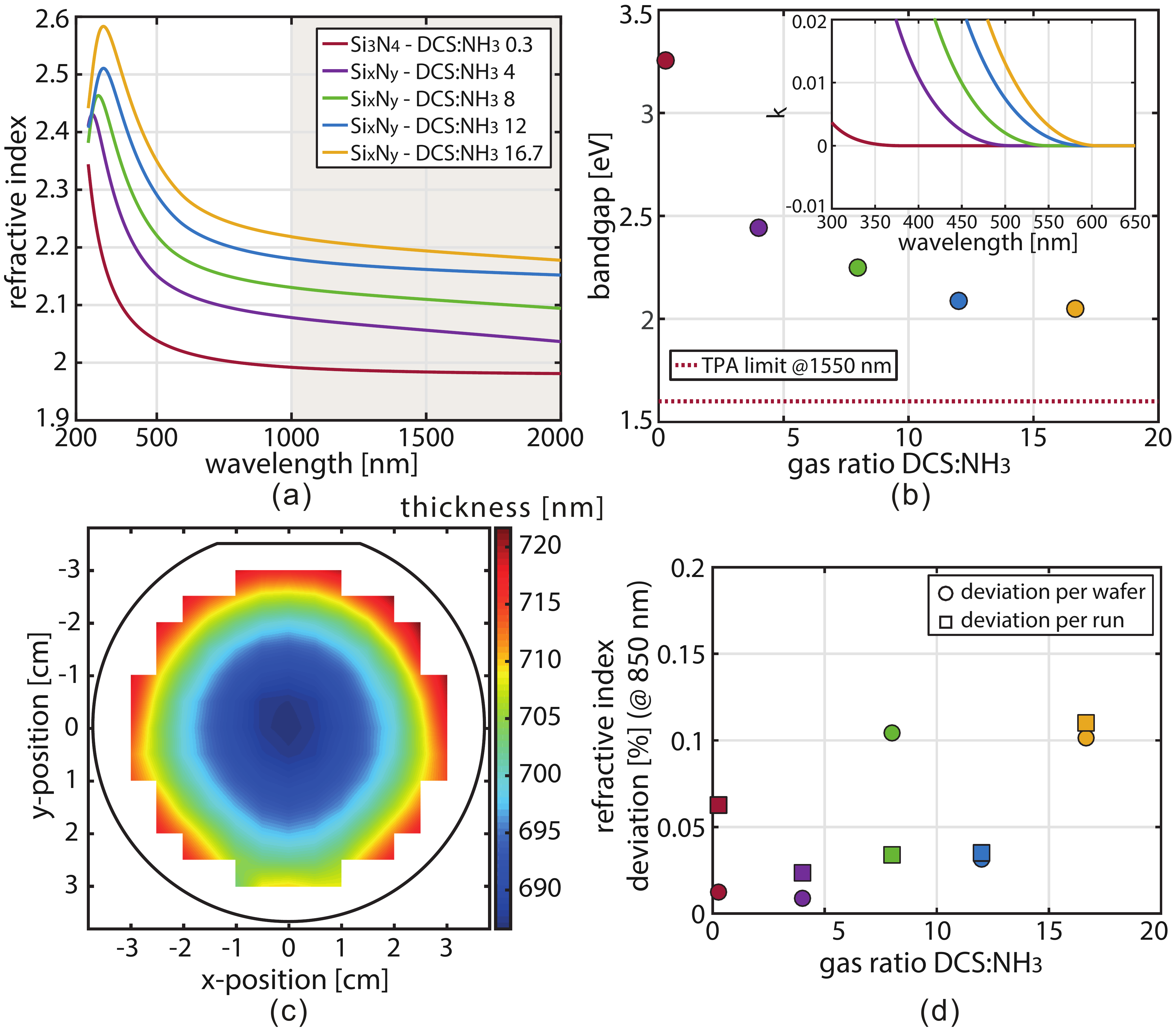}
	\caption{a) Measured refractive index as a function of wavelength for different SiN compositions. The shadowed area shows extrapolated data from the measurements. b) Optical bandgap as a function of gas flow ratio. The two-photon absorption limit indicates twice the photon energy at 1550 nm wavelength. The inset shows a zoomed in part of the measured imaginary refractive index $k$ as a function of wavelength for the fabricated compositions.  c) Area scan of the film thickness with 139 points showing the uniformity of the deposited SiN film on a 3-inch wafer (DCS:NH$_\textrm{3}$ 8). d) Reproducibility of the refractive index at 850 nm carried out over 22 points on a single wafer and between 3 independent deposition runs.}
	\label{fig1}
	\end{centering}
	\end{figure}

We fabricated high-confinement strip waveguides with a silicon nitride core surrounded by a silicon dioxide cladding. For the deposition of silicon nitride (SiN), we used dichlorosilane (DCS) and ammonia (NH$_3$) as precursor gases in an LPCVD furnace (Centrotherm). In order to form the waveguide core, the SiN film was patterned by optical contact lithography (DUV) and dry etching (CHF$_3$+O$_2$). The process steps were similar to the ones presented in \cite{kruckel:Opt-Exp} and further advanced by including a standard cleaning step after etching.\\
Throughout the fabrication and characterization process, it became clear that the ratio between the two precursor gases DCS and NH$_3$ is the most relevant parameter to control the properties of the SiN film. The ratio DCS:NH$_3$ directly impacts the atomic composition of the material. It yields an increase in silicon content with increased gas ratio \cite{refractive_index_vs_siN2}. We have confirmed this trend with measurements based on X-ray diffraction \cite{kruckel:Opt-Exp}. Throughout this paper we will however use the ratio DCS:NH$_3$ to describe the composition because it provides a more accurate means to control the fabrication process. The relation between gas flow ratio and atomic composition in LPCVD SiN has been reported in the literature \cite{refractive_index_vs_siN2}.\\
We fabricated five different SiN compositions. The chosen gas ratios DCS:NH$_3$ reached from 0.3 for stoichiometric silicon nitride (Si$_3$N$_4$) to non-stoichiometric silicon nitride (Si$_{\mathrm{x}}$N$_{\mathrm{y}}$) with a maximum gas ratio of 16.7 given by the gas flow limits of the furnace. For deposition of Si$_3$N$_4$ we used the recipe provided by Centrotherm. All films were deposited under similar temperature conditions around 800$^{\circ}$C and the pressure was optimized to reach both stable deposition conditions and best possible film uniformity.\\
The target height and width of the fabricated waveguide core were 700 by 1650 nm for all compositions. Stoichiometric Si$_3$N$_4$ films display cracks at this thickness, so we etched crack barriers following a similar process to \cite{contr-cracking,crack-barrier-lipson}  into the bottom oxide using a laser writer tool followed by a wet-etch step in order to minimize waveguide imperfections. The patterned structures increased the crack-free area of the wafer to several cm$^2$ as cracks originating from wafer handling were stopped. However, full control of cracks was not achieved. Etching into the silicon substrate was suggested in \cite{crack-barrier} to get rid of this issue. It is worth emphasizing that films deposited using gas ratios above $\sim$4 had no cracks at the fabricated thickness of 700 nm owing to a decrease in stress for silicon-rich nitride. As an example, the measured stress in stoichiometric Si$_3$N$_4$ films (DCS:NH$_3$ 0.3) lied above 1200 MPa whereas films from DCS:NH$_3$ 12 had values around $\sim$500 MPa.\\
\noindent We used spectroscopic ellipsometry for the analysis of the refractive index properties of the materials. We fitted the measurement data using a Tauc-Lorentz oscillator model \cite{tauc-lorentz}. This model yields the resonance pole information and optical bandgap. Figure \ref{fig1}(a) shows the real part of the refractive index for all fabricated compositions within the wavelength range of the ellipsometer (245-1000 nm) and the extrapolation up to 2000 nm.  One can observe that, at a fixed wavelength, the refractive index increases with higher gas flow ratio. This is consistent with an increase of silicon content in the film \cite{refractive_index_vs_siN2}. Figure \ref{fig1}(b) illustrates the variation of the optical bandgap of the material with the gas flow ratio. It is clear that by increasing the gas flow ratio, the optical bandgap of the film decreases, but it always remains above the energy level for two-photon absorption to take place at 1550 nm, as indicated by the dotted red line. Physically the measured optical bandgap corresponds to the photon energy at which the imaginary part of the refractive index, $k$, becomes non-zero, as shown in the figure inset. The absence of material loss for wavelengths above 350-600 nm clearly indicates the ultra-broad transparency window of silicon nitride. \\
We studied the uniformity of the SiN films by realizing repeated ellipsometer area scans. The uniformity was measured with 137 points across the 3-inch wafer with an x-y-position change of 0.5 cm. For illustration purposes we used a SiN film fabricated with the gas ratio 8 in Fig. \ref{fig1}(c), but similar results were obtained for the other compositions and over independent fabrication runs. A maximum deviation from target (700 nm) below 3\% can be observed. Such non-local change in film height should be considered when designing waveguide dimensions for dispersion engineering. Next, we studied the reproducibility of the film composition by first comparing the refractive index change across a single wafer (22 measurement points) and second by comparing the average refractive index of 3 wafers from independent deposition runs. The results are shown in Fig. \ref{fig1}(d). A deviation below 0.15\% is reached in both cases when measuring $\sim$200 nm thick films. Thin films were chosen here to reduce processing time. This study shows that the gas flow ratio in an LCPVD furnace provides a reliable way to control the optical properties and composition of the SiN films.\\

\section{Linear waveguide properties}

%%%%%%%%%%%%%%%%%%%%%%%%%%%%%%%%%%%%%%%%%%%%%%%%%%%%%
%%	
\subsection{Waveguide propagation loss}
%%
%%%%%%%%%%%%%%%%%%%%%%%%%%%%%%%%%%%%%%%%%%%%%%%%%%%%%

\begin{figure*}[!b]
  \begin{centering}
  \includegraphics[width=0.99\textwidth]{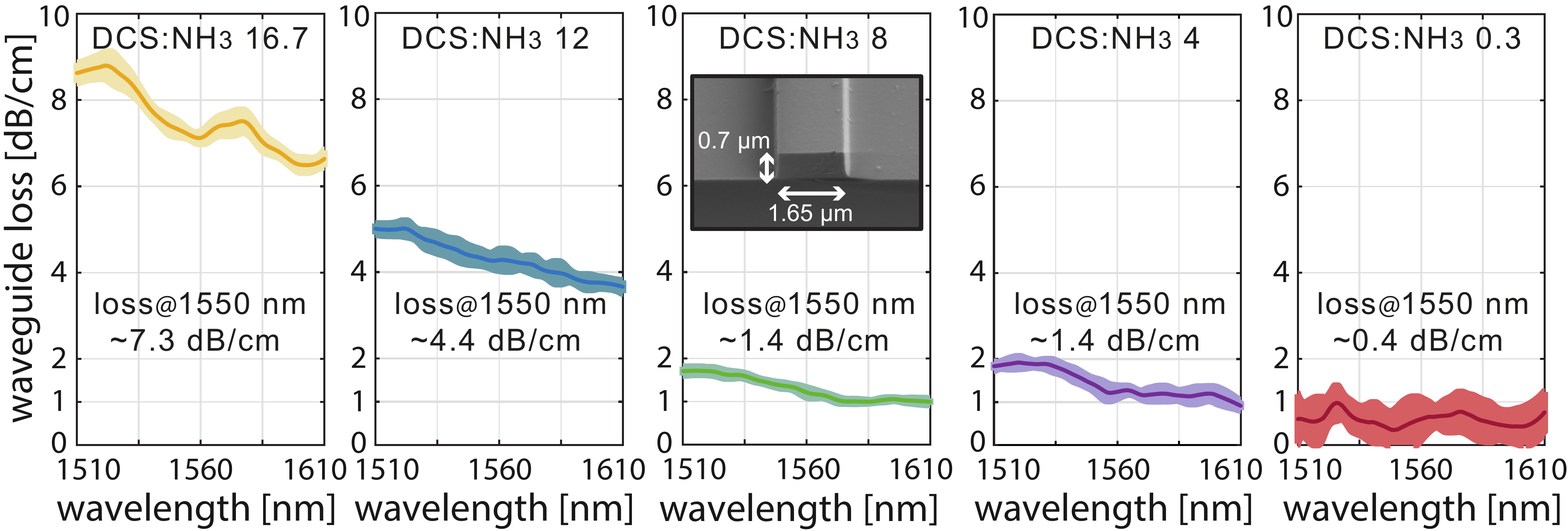}
	\caption{Waveguide propagation loss as a function of wavelength for waveguides with 5 different silicon nitride compositions corresponding to different DCS:NH$_3$ ratios. The darker colored line shows the mean value and the brighter shadowed areas illustrates the standard deviation of the loss measurements. The inset shows the waveguide core after etching.}
	\label{fig2}
	\end{centering}
	\end{figure*}

We evaluated waveguide propagation losses and coupling losses using the cut-back method with 3 different lengths. The loss measurements were done in a spectrally resolved manner from 1510 to 1610 nm and the polarization was optimized for maximum system throughput. In order to present representative data, we performed the measurements over 5 waveguides. The averaged values and standard deviation per wavelength are presented in Fig. \ref{fig2}. Since we used the same cross-section geometry for all waveguide compositions, any change in measured waveguide propagation loss is likely caused by the material properties of the SiN waveguide core.\\
For the waveguides with gas flow ratio 4 to 16.7 the three cut-back lengths were 4, 3 and 1 cm. The waveguides fabricated from Si$_3$N$_4$ had lengths of 2.2, 1.3 and 0.9 cm because we had to restrict ourselves to the crack-free area of the wafer. Instead of the straight waveguides used in this work, a space saving design (spiral, meander) is recommended if film cracks cannot be avoided. We used horizontal coupling directly into the waveguide core via tapered lensed fibers with 2.5 $\micro$m spot size diameter. Coupling losses $\sim$4 dB per facet were obtained for all compositions. The larger uncertainty in the Si$_3$N$_4$ waveguide losses is due to the shorter cut-back lengths, resulting in lower measurement precision. We observe a reduction in waveguide loss with decreasing gas flow ratio. The typical absorption resonance around 1.53 $\micro$m is visible for all compositions with DCS:NH$_3$ $\ge$4. This is likely due to the existence of N-H bonds that remain from the use of ammonia as precursor gas. This fact alone however cannot explain the increase absorption observed in the L band. Interestingly, we could get rid of the absorption resonance for DCS:NH$_3$ 0.3 by realizing an annealing step at 1200$^{\circ}$C over 3 h in nitrogen atmosphere. The final losses after annealing for Si$_3$N$_4$ were around 0.4 dB/cm at 1550 nm wavelength, a value comparable to other reported values in the literature using waveguides with similar dimensions \cite{lionix-sc-gen}. We estimate the scattering losses contribute 0.2 dB/cm \cite{kruckel:Opt-Exp}. Additional steps, such as the use of a higher quality LPCVD top oxide layer could further improve the propagation losses \cite{crack-barrier}. How well the annealing step can be applied to the other composition requires further investigations. For example the same annealing recipe caused an increase in waveguide loss for the silicon-rich composition DCS:NH$_3$ 8. We suggest that the losses are caused by the formation of silicon clusters, since a similar effect has been reported in the literature when annealing silicon-rich nitride films \cite{annealing_sirich}. Whether the losses of the silicon-rich nitride films can be decreased with a different annealing recipe deserves further investigation because, as we shall show in the next section, these materials display very high nonlinear Kerr coefficient.\\

%%%%%%%%%%%%%%%%%%%%%%%%%%%%%%%%%%%%%%%%%%%%%%%%%%%%%
%%
\subsection{Numerical analysis of waveguide dispersion}
%%
%%%%%%%%%%%%%%%%%%%%%%%%%%%%%%%%%%%%%%%%%%%%%%%%%%%%%
\begin{figure*}[!b]
  \centering
	\includegraphics[width=1.0\linewidth]{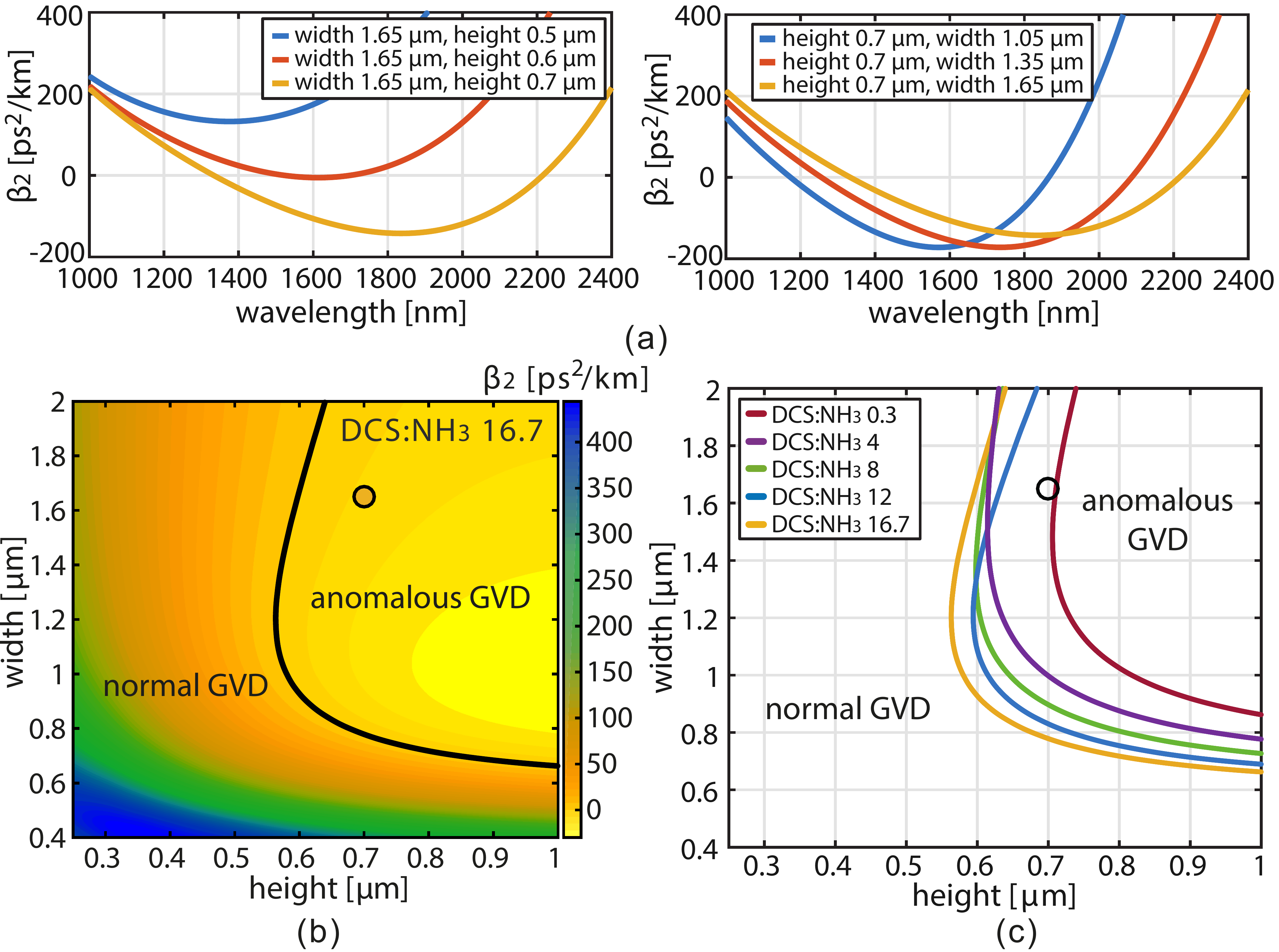}
	\caption{a) Impact of height and width variation on the group velocity dispersion (DCS:NH$_3$ 16.7, quasi-TE-mode). b) Group velocity dispersion coefficient $\beta_2$ as a function of waveguide width and height of the silicon nitride waveguide considering the material corresponding to DCS:NH$_3$ 16.7 (1550 nm wavelength, quasi-TE-mode). The black line indicates the dimensions at which zero GVD occurs. The circle indicates the dimensions of the fabricated waveguides.  c) Waveguide dimensions at which crossing from normal to anomalous dispersion occurs (1550 nm wavelength, quasi-TE-mode). The circle indicates the dimensions of the fabricated waveguides.  }
	\label{fig3}
	\end{figure*}

Here we look into the dispersion engineering possibilities of the SiN waveguides. The waveguide dispersion can be engineered over a broad wavelength range by tailoring its dimensions \cite{foster-broad-band} or in the case of a silicon nitride waveguide by changing its material composition \cite{ihp-sin}.\\
The simulations were performed with a modesolver based on COMSOL-Multiphysics using the refractive index data for the core and cladding materials extracted from ellipsometer measurements. First, we show how the waveguide width and height individually impact the group velocity dispersion profile $\beta_2$. This is exemplified in Fig. \ref{fig3}(a) for the SiN composition fabricated using the gas ratio 16.7 for the fundamental quasi TE-mode. The left illustration shows that an increase in waveguide height is required (fixed width of 1.65 $\micro$m) in order to reach anomalous GVD. In the right illustration it can be seen that varying the waveguide width allows to shift the second zero dispersion wavelength to lower optical frequencies.\\
The impact of both waveguide width and height together is studied in Fig. \ref{fig3}(b) in more detail. Here, we show how the GVD changes for all geometry combinations for a width from 0.4 to 2 $\micro$m and a height from 0.25 to 1 $\micro$m. This allows relating the dispersion profile to the optical confinement. The illustration is valid for the fundamental quasi-TE-mode at wavelength of 1550 nm. Waveguide dimensions that lead to $\beta_2$ values of zero are lying along the black line. Thus, anomalous dispersion is achieved for the dimensions on the right side of this line. We compared the dimensions required to cross from normal to anomalous dispersion for all fabricated SiN compositions, see Fig. \ref{fig3}(c). The larger refractive index of the silicon rich compositions causes in general a shift of the anomalous GVD crossing towards thinner and narrower dimensions. This trend is expected, since a higher silicon rich composition provides higher refractive index contrast and allows for higher mode confinement in the core. This result will be further discussed in the next subsection in the context of optimization of the nonlinear parameter. It can be seen that for the fabricated dimensions, anomalous dispersion at 1550 nm is achieved for all compositions except for Si$_3$N$_4$.

%%%%%%%%%%%%%%%%%%%%%%%%%%%%%%%%%%%%%%%%%%%%%%%%%%%%%
%%	
\section{Nonlinear properties}
%%
%%%%%%%%%%%%%%%%%%%%%%%%%%%%%%%%%%%%%%%%%%%%%%%%%%%%%

\begin{figure}[!b]
  \centering
	\includegraphics[width=1.0\linewidth]{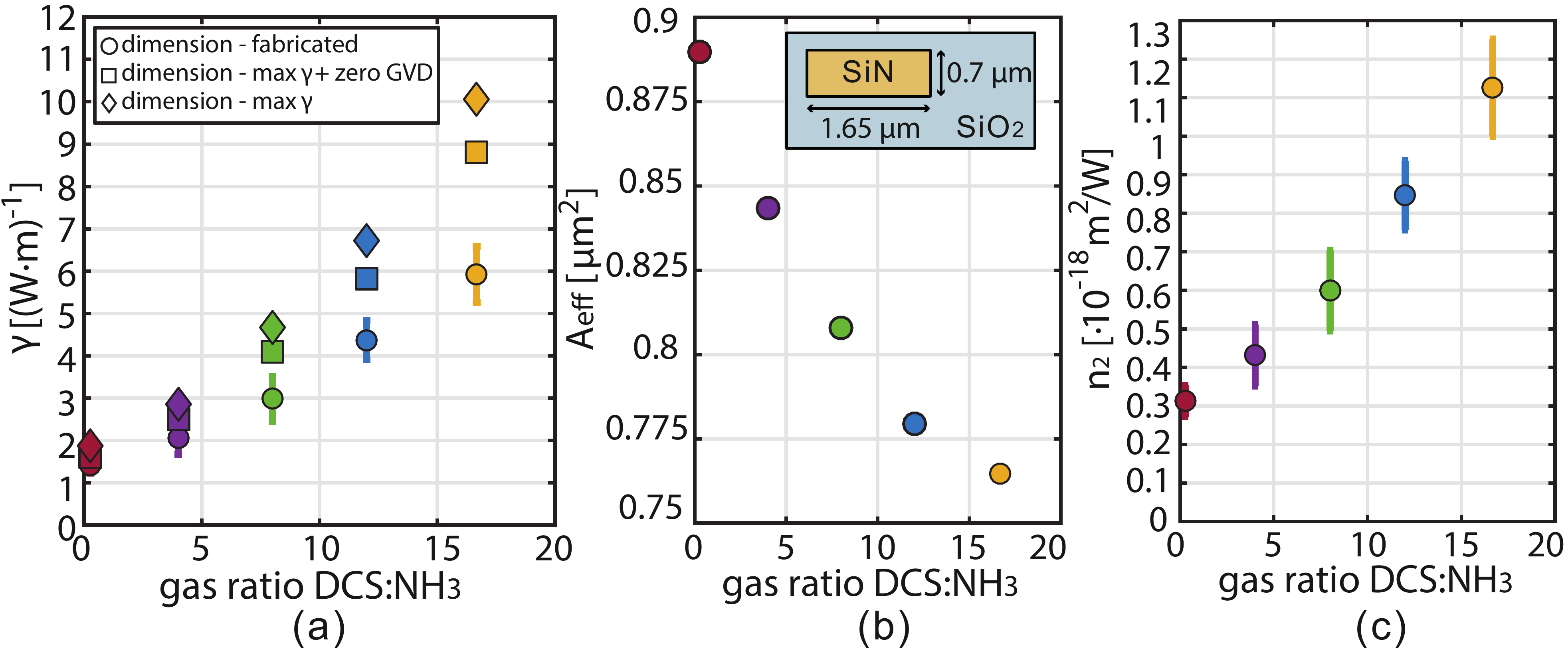}
	\caption{a) Circles - Measured nonlinear parameter $\gamma$ with the fabricated dimensions (0.7 $\micro$m height and 1.65 $\micro$m width). Diamonds - Numerical simulations of the maximum $\gamma$ at the dimensions that give the highest optical confinement in the waveguide. Squares - Numerical simulations of the largest achievable $\gamma$ with the restriction to have zero group-velocity dispersion. b) Simulation results of the effective area $A_{\mathrm{eff}}$ for the fabrication gas ratios. c) Evaluated nonlinear Kerr coefficient from the measured nonlinear parameter $\gamma$ and the simulated effective area $A_{\mathrm{eff}}$.}
	\label{fig4}
	\end{figure}
	
	\begin{figure}[!t]
  \centering
	\includegraphics[width=1.0\linewidth]{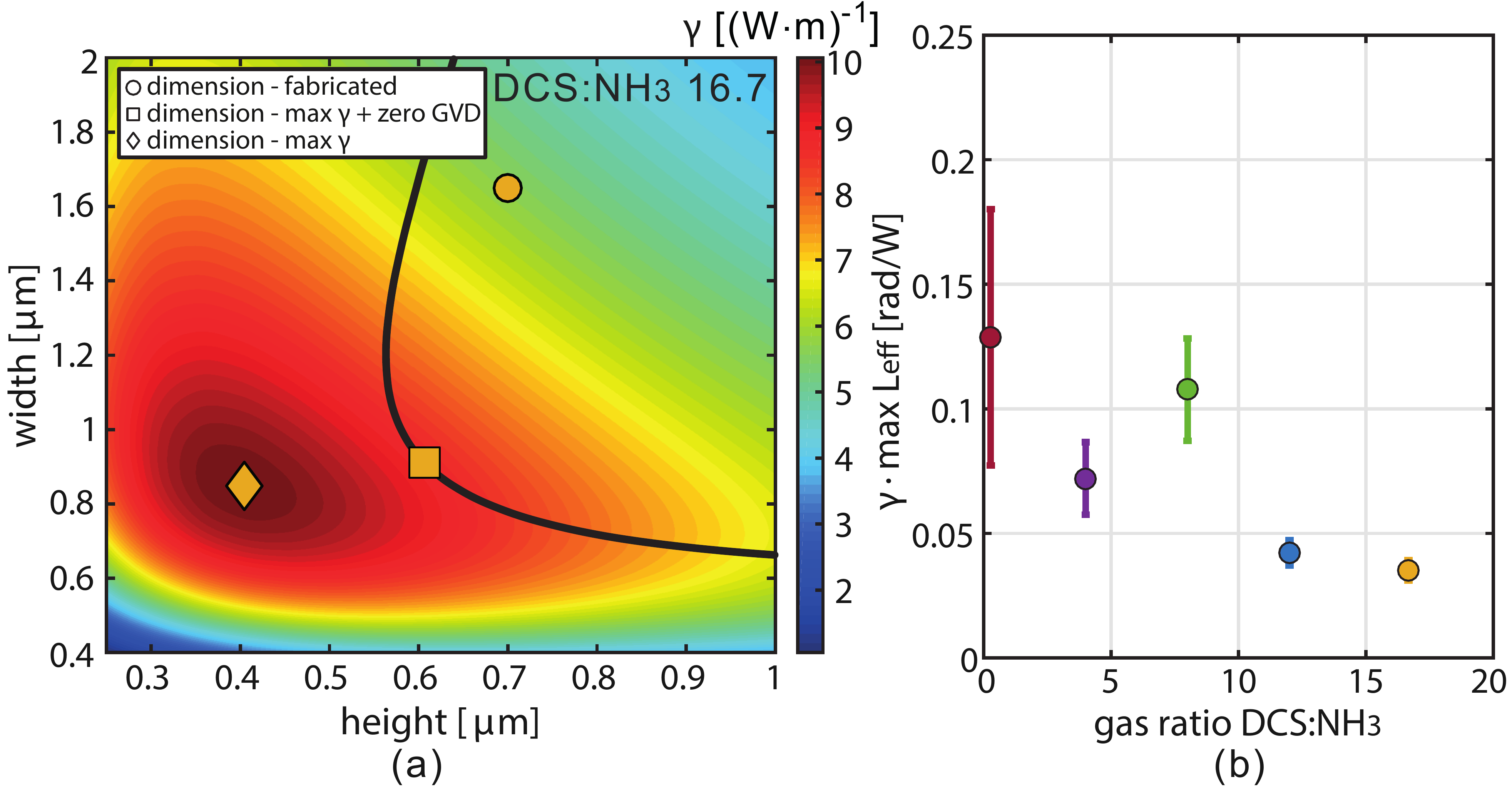}
	\caption{a) Nonlinear parameter for different width and height of the waveguide core fabricated from DCS:NH$_3$ ratio 16.7. The simulations are made for the fundamental quasi TE-mode at 1550 nm wavelength. The plot indicates the fabricated waveguide dimensions, the dimensions that lead to a maximum nonlinear parameter, and the dimensions that result in a maximum nonlinear parameter with the requirement of zero GVD. b) Maximum relative nonlinear phase shift $\gamma\cdot \mathrm{max}\,L_{\mathrm{eff}}$ achieved for the fabricated waveguide dimensions of 700 nm height and 1650 nm width.}
	\label{fig5}
	\end{figure}	
	
In this section we illustrate that the nonlinear Kerr coefficient of silicon nitride strongly depends on the gas flow ratio utilized during deposition. First, we evaluate the nonlinear parameter $\gamma$ in measurements based on a dual-pump experiment \cite{dual-pump,kruckel:Opt-Exp} where the nonlinear phase shift is measured as a function of optical pump power, i.e. $\theta_{\mathrm{nl}}=\gamma PL_{\mathrm{eff}}$, with $L_{\mathrm{eff}}$ being the effective length of the waveguide. The measurements were carried out around 1563 nm. This experiment also served as a means to verify the absence of multi-photon absorption at telecom wavelengths up to several hundreds of mW continuous-wave (CW) pump power coupled into the waveguide. The slope in the linear fit of $\theta_{\mathrm{nl}}$ vs $P$ provides an estimation of the nonlinear parameter $\gamma$ of the waveguide with knowledge of $L_{\mathrm{eff}}$. The results of the measurements are summarized in Fig. \ref{fig4}(a) by the circular symbols. The calculation of the errors includes both the error in the slope and the error in $L_{\mathrm{eff}}$ provided by the uncertainty in the losses (Fig. \ref{fig2}). It is clear that increasing the flow ratio yields a higher nonlinear parameter. This however does not necessarily mean that the nonlinear Kerr coefficient increases. To verify this, we calculated the Kerr coefficient as $n_2 = \gamma \lambda A_{\mathrm{eff}}/2\pi$. The effective area can change even if the cross-section is identical for all compositions. This fact is illustrated in Fig. \ref{fig4}(b). The effective area decreases for larger DCS:NH$_3$ ratios, which results from the higher optical confinement owing to the increase in refractive index. Figure \ref{fig4}(c) shows the estimated $n_2$ for each of the measured SiN compositions. It is clear that the Kerr coefficient can be modified by roughly a factor of 4 by controlling the gas flow ratio during film deposition; from $n_2 = (0.31 \pm 0.04) \cdot 10^{-18}$ m$^2$/W for Si$_3$N$_4$ (DCS:NH$_3$ 0.3) to $n_2 = (1.13 \pm 0.13) \cdot 10^{-18}$ m$^2$/W for DCS:NH$_3$ 16.7.\\
The results in Fig. \ref{fig4}(a) (circles) correspond to the nonlinear Kerr parameter measured for different compositions with identical cross-section. This parameter could be further enhanced by optimizing the waveguide dimensions for each composition. This is illustrated in Fig. \ref{fig5}(a) using the composition corresponding to DCS:NH$_3$ 16.7. The parameter $\gamma$ is maximized for a waveguide whose dimensions provide the highest optical confinement. The dimensions correspond to 0.82 $\micro$m width and 0.4 $\micro$m height (highlighted by the diamond symbol in Fig. \ref{fig5}(a)). This maximum value is almost double the value we measured for the fabricated geometry. However, comparing this plot with Fig. \ref{fig3}(b), it becomes clear that this cross-section geometry does not yield anomalous dispersion. This conclusion is generally valid for all compositions. One could instead calculate the geometry that provides the maximum nonlinear parameter within the zero GVD contour in Fig. \ref{fig3}(b) (plotted again in Fig. \ref{fig5}(a)). This point is defined by the square symbol in Fig. \ref{fig5}(a). This analysis is carried out for all compositions and the results are presented in Fig. \ref{fig4}(a), where the measured nonlinear parameter for each composition is compared to what could be obtained if the waveguide dimensions were optimized. The nonlinear parameter could be enhanced by almost an order of magnitude by varying the stoichiometry of the film with respect to the measured value in stoichiometric silicon nitride.\\
Of course, the relevant parameter for most nonlinear optics applications in absence of multiphoton absorption is the amount of nonlinear phase shift per unit power, i.e. $\theta_{\mathrm{nl}}/P=\gamma L_{\mathrm{eff}}$. This figure of merit is important, as the nonlinear parameter alone does not provide the whole picture. In order to include propagation losses, we considered the case of maximum achievable effective length, max $L_{\mathrm{eff}}$,  as $1/\alpha$ with $\alpha$ being the linear loss. The results for our materials are shown in Fig. \ref{fig5}(b). Surprisingly, the plot indicates two optimum compositions that yield maximum nonlinear phase shift. Composition DCS:NH$_3$ 0.3 is stoichiometric silicon nitride and it benefits from very low propagation losses. Composition DCS:NH$_3$ 8 corresponds to the results previously reported in \cite{kruckel:Opt-Exp,sc_dtu}. This material reaches a similar nonlinear phase shift due to its increased nonlinear coefficient and moderate linear loss. The large uncertainty in the standard deviation for the Si$_3$N$_4$ material comes from the high uncertainty of the low waveguide losses, which has a larger impact when calculating max $L_{\mathrm{eff}}$. It is important to emphasize that this comparison is not fundamental. Further decrease in propagation losses for one or another composition could dramatically change this picture.

%%%%%%%%%%%%%%%%%%%%%%%%%%%%%%%%%%%%%%%%%%%%%%%%%%%%%
%%
\section{Theoretical comparison and discussion}
%%
%%%%%%%%%%%%%%%%%%%%%%%%%%%%%%%%%%%%%%%%%%%%%%%%%%%%%

\begin{figure}[!b]
  \centering
	\includegraphics[width=1.0 \linewidth]{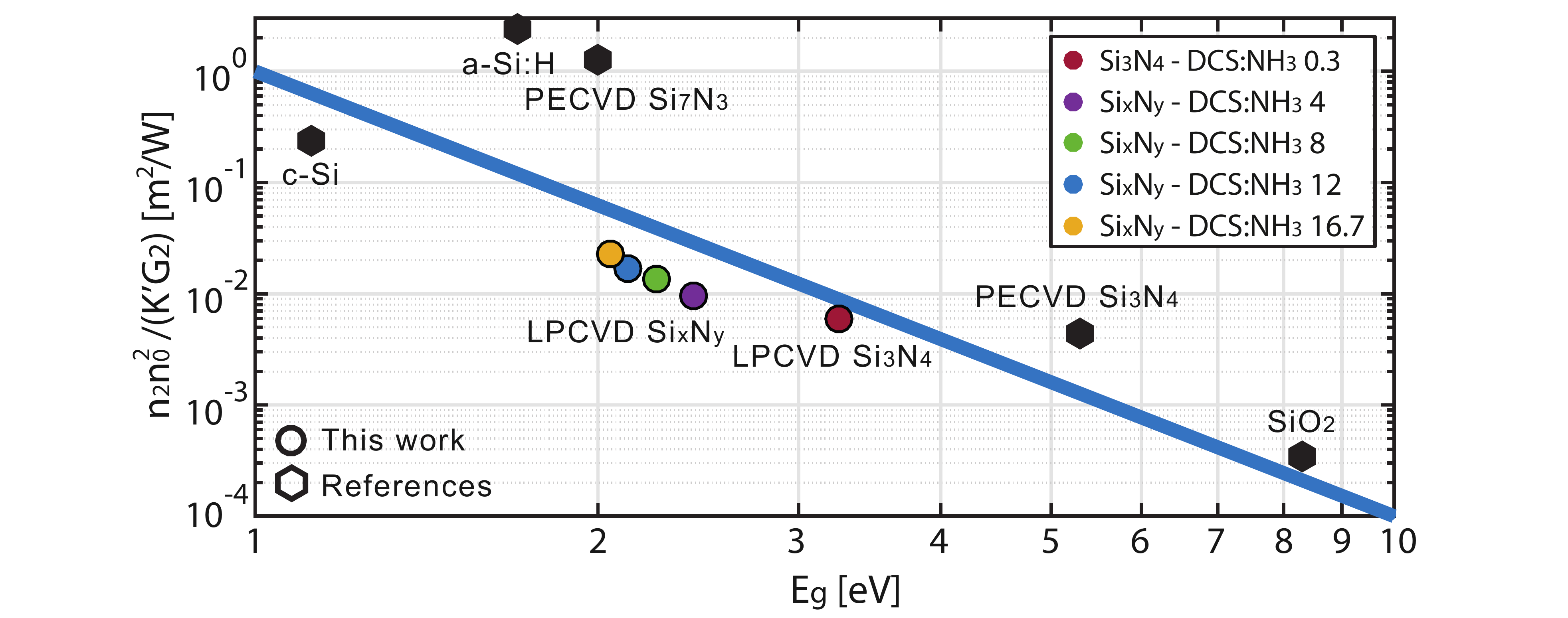}
	\caption{Rescaled nonlinear Kerr coefficient as a function of optical bandgap. The blue line indicates the $1/E_g^4$ relation. See text for more details.}
	\label{fig6}
	\end{figure}

In order to understand the fundamental limitations of the Kerr effect in our platform and to validate our experimental results, we compare our measured values of the nonlinear Kerr coefficient $n_{2}$ with theoretical expectations. A useful theoretical approach developed by Sheik-Bahae et al. \cite{sheik-bahae} links $n_{2}$ directly to the optical bandgap of the material according to 
\begin{equation}
n_2 =K'\frac{G_2(\hbar\omega/E_\mathrm{g})}{n_0^2E_\mathrm{g}^4}.
\label{sheik-bahae}
\end{equation}
This equation is developed under the assumption that there is a Kramers-Kronig relation between the real and imaginary parts of the nonlinear refractive index and that the nonlinear absorption is dominated by two-photon absorption \cite{sheik-bahae}. The relation is valid for a broad range of photon frequencies, even when they are far below the optical bandgap. This is taken into account by the function $G_2$. In the equation above, $n_0$ is the refractive index and $K'$ is a constant. The function $G_2$ is maximized for photon energies close to $E_\mathrm{g}/2$ thus suggesting that highest Kerr nonlinearities are attained when using a wavelength close to the TPA limit. The formulation is valid for solids provided the assumptions above hold. In such case, it gives reasonable estimates for a broad range of oxides and glasses \cite{tanaka}.\\
\noindent Figure \ref{fig6} indicates that the properties of our materials follow reasonably well the Sheik-Bahae relationship. The photon energy and refractive index were taken at 1563 nm, the wavelength of the dual-pump experiment. Uncertainties from the Kerr coefficient evaluation are not considered here. For comparison, other widely used nonlinear materials are included in Fig. \ref{fig6} based on data of refractive index, nonlinear Kerr coefficient and bandgap provided in the references. The way the bandgap is defined is not explicitly mentioned in the publications. This may lead to a discrepancy in $E_\mathrm{g}$. At the two extremes of the bandgap scale, we find crystalline silicon (c-Si) \cite{c-Si-leuthold,Si_Eg} and silicon dioxide (SiO$_2$) \cite{Sio2_Eg,Sio2_n2}. These materials display different Kerr nonlinearities in comparison to our silicon nitride compositions but they follow qualitatively the scaling $1/E_\mathrm{g}^4$. It is interesting to note that the two low bandgap materials deposited with PECVD, namely hydrogenated amorphous silicon (a-Si:H) \cite{a-Si-h,a-Si-h-kuyken} (refractive index of 3.5 was assumed) and Si$_7$N$_3$ \cite{PECVD-SiN-singapore} display a huge increase in nonlinearities but a significant deviation from the Sheik-Bahae relation. Our LPCVD Si$_3$N$_4$ display a Kerr coefficient value [$n_2 = (0.31 \pm 0.04) \cdot 10^{-18}$ m$^2$/W] in agreement with previously reported values in the literature \cite{tien}. This value is also compatible with the value reported for PECVD Si$_3$N$_4$ \cite{ikeda,tan_ikeda}.

%%%%%%%%%%%%%%%%%%%%%%%%%%%%%%%%%%%%%%%%%%%%%%%%%%%%%
%%
\section{Conclusion}
%%
%%%%%%%%%%%%%%%%%%%%%%%%%%%%%%%%%%%%%%%%%%%%%%%%%%%%%

We have demonstrated that the gas flow ratio in the LPCVD process to synthesize silicon nitride can be used to control the nonlinear characteristics of an integrated waveguide. Essentially, the gas flow ratio modifies the frequency location of the optical bandgap, which in turn varies the refractive index and Kerr coefficient. These parameters increase with higher gas flow ratios. We did not observe nonlinear absorption in the erbium telecommunications band up to hundreds of mW of optical power coupled into the waveguide. The nonlinear Kerr coefficient of LPCVD silicon rich nitride follows qualitatively a Sheik-Bahae relationship. We found two compositions yielding an optimum nonlinear phase shift per launched power, one was stoichiometric silicon nitride (Si$_3$N$_4$) and the other one a composition with slightly increased silicon content \cite{kruckel:Opt-Exp}. The performance of the former is due to the relatively low linear losses, whereas the latter provides an increase in nonlinear parameter. The possibility to modify the refractive index of the waveguides in a highly reproducible fashion opens up a new way to engineer the dispersion of silicon nitride waveguides and enhance their nonlinear characteristics.

\section*{Funding}
This work was supported by the European Council under grant agreement ERC-2011-AdG-291618 PSOPA, the Wallenberg Foundation and the Swedish Research Council (VR).

\end{document}